\documentstyle[psfig,11pt]{article}

\def\la{\hbox{\rlap{\raise.3ex\hbox{$<$}}\lower.8ex\hbox{$\sim$}\ }}
\def\ga{\hbox{\rlap{\raise.3ex\hbox{$>$}}\lower.8ex\hbox{$\sim$}\ }}
\def\deg{$^{\circ}$}
\def\ktbb{kT$_{\rm BB}$}
\def\rbb{R$_{\rm BB}$}

\def\rxte{{\it Rossi X-ray Timing Explorer (RXTE)}}

\def\chisqr{$\chi^{2}_{red}$}
\def\onee{1E1724-3045}
\def\tz{1E1724-3045}
\def\ergs{ergs s$^{-1}$}

\def\nhv{$\times 10^{22}$ cm$^{-2}$}
\def\nh{N$_{\rm H}$}

\def\etal{{\it et al.\/}}
\def\kte{kT$_e$}
\def\ktw{kT$_W$}

\begin{document}
\vspace{1.0cm}

{\Large \bf RXTE Broad Band X-ray Spectrum of the Burster 1E1724-3045}

\vspace{1.0cm}

J. F. Olive$^1$, D. Barret$^1$, L. Boirin$^1$ and
J. E. Grindlay$^2$

\vspace{1.0cm}

$^1${\it CESR, CNRS/UPS, 9 Avenue du Colonel Roche, 31028 Toulouse Cedex 04, France (olive@cesr.fr)}\\ 
$^2${\it Harvard Smithsonian Center for Astrophysics, 60 Garden
Street, Cambridge, MA 02138, USA} \\

\vspace{0.5cm}

\section*{ABSTRACT}

The X-ray burster \tz~located in the globular cluster Terzan 2 is
known as one of the persistent (though variable) hard X-ray sources as
shown by the SIGMA observations of the Galactic Center region. \tz~was
observed with the PCA and HEXTE experiments onboard the \rxte~on
November 1996 for about 100 ksec. The broad band spectral capability
and sensitivity of RXTE enables to study simultaneously the X-ray
(3-20 keV) and hard X-ray components (E$>$20 keV) of this
source. During the observation, this ``Atoll'' source was in its
``Island'' state characterized by a hard Comptomized spectrum with an
electron temperature \kte $\sim 29$ keV, an optical depth $\tau \sim
2.9$ (spherical geometry) and a temperature of the ``seed'' photons of
\ktw $\sim 1.2$ keV. Below 5 keV, there is a soft excess which we fit with
a blackbody of \ktbb $\sim 0.67$ keV. The Comptonization temperature
is significantly lower those observed for black holes candidates in
their low luminosity state (\kte~$\ga$~50 keV).  Finally, our
observation allows us to associate the presence of an hard tail with a
low luminosity X-ray state (1-20 keV luminosity of 1.0 $\times
10^{37}$ \ergs~at 7.7 kpc).

\section{INTRODUCTION} 

It is now well known that, like black hole binaries, X-ray bursters
(XRB, hence neutron star binary systems) can emit hard X-rays (see
Tavani and Barret, 1997 and references therein). During the last few
years, a dozen of such sources have been detected, mostly with the
SIGMA and BATSE instruments. To investigate the conditions under which
a neutron star can emit hard X-rays and the residual differences
between black holes and neutron star systems (such as the position of
the energy cutoff and luminosity) a high sensitivity broad band
spectrum from X-rays to hard X-rays is necessary. Indeed, the hard
X-ray data alone are inadequate to discriminate between thermal and
non-thermal models. Furthermore, due to the moderate sensitivity of
hard X-ray telescopes such as SIGMA and BATSE, only mean spectra,
averaged over at least few weeks, are reliable.

The situation changed recently with experiments such as RXTE. This
experiment allows to study these systems with a good sensitivity in a
broad energy band (from 3 to 200 keV) by combining of the PCA and
HEXTE instruments (Bradt \etal, 1993). One of the most natural target
for this study is \tz. This source was one of the first X-ray burster
detected above 100 keV. The detection came with SIGMA in the course
its first observation of the Galactic Center region in March-April
1990 (Barret \etal, 1991). At that time, this source was the third
brightest hard X-ray source of the GC field (behind 1E1740-2942 and
GRS1758-258) and was characterized by a very hard power law spectrum
(photon index of $1.7\pm0.5$) extending up to 300 keV (Barret \etal,
1991). Subsequently, over more than five years of GC observation by
SIGMA, \tz~was continuously detected with a flux ranging from 10 to 50
mCrab in the 35-75 keV band and a mean flux of $\sim 22$ mCrab
(Goldwurm \etal, 1995 and Churazov \etal, 1997). The very hard
spectrum of the first observation was never observed again. The
time-averaged hard X-ray spectrum of the source can be approximatively
fitted by a power law with a photon index of $3.0\pm0.3$, and is
probably softer if the data of the March-April 1990 observation are
excluded ($\alpha=3.3\pm0.4$, see Goldwurm \etal,
1995). Alternatively, this spectrum could also be fit with a Thermal
Bremsstrahlung model of temperature $\sim 50$ keV (Churazov
\etal, 1997).

In the classical X-ray range (1-20 keV), the source has been
repeatedly observed by various X-ray experiments since its discovery
in 1977 by UHURU (Swank \etal, 1977). The most reliable spectral
observations of \tz~(i.e. those performed with EXOSAT and TTM) showed
that the source is characterized by a relatively hard power law type
spectrum in X-rays (photon index of $\sim 2.0$). These observations,
combined with the mean SIGMA spectrum suggested that a spectral break
occurs somewhere below 100 keV. More recent observations with ASCA
(Barret \etal, 1998) and SAX LECS-MECS (Guainazzi \etal, 1998) showed
that the source spectrum can be fit with a Comptonized model (\kte
$\sim 30.0$ keV) and a soft blackbody component (\ktbb $\sim 0.6$
keV).

In this paper, we report on the joint spectral analysis of the PCA
and HEXTE data on \onee. In the last section, we make a brief
comparison with black hole candidates. For a full description of the
observation and the timing analysis of the source variability, we
refer to Olive \etal~(1998).

\section{The RXTE observation and data analysis} 

The RXTE observation took place on November, 4th to 8th, 1996 for a
total exposure time of about 100 kiloseconds. The data have been
filtered out for elevation greater than 10\deg~(as recommended) and
source pointing offset less than 0.02\deg. We have also excluded data
recorded during the SAA passage (and 30 minutes after it) using the
FTOOLS 4.1 version of xtefilt.

\subsection{The PCA data}

First, using the PCA {\it standard 2} data, we have built light
curves, color-color, and hardness-intensity diagrams with various time
resolution. The FTOOLS used for PCA background estimation was the
latest pcabacest version 2.0c. Along the RXTE observation, no time or
spectral variability could be inferred from this analysis. So we
decided, for spectral analysis, to consider the mean spectra over the
whole observation. The spectra of each PCA unit (PCU0 to PCU4) have
been extracted separately and PCU response matrices have been made for
each PCU using the FTOOLS pcarsp 2.36.  Then we have combined the 5
PCU data and matrices using the FTOOLS addspec 1.2.0.  In order to
investigate the possible systematics in the PCA data we have first
analyzed a 15 ksec Crab observation (March 22, 1997, similar modes of
observation). Provided that $1\%$ systematics are added to the data,
the Crab spectrum was correctly fitted, so we decided to add, before
fitting the \tz~data, the same $1\%$ systematics to the data.

\subsection{The HEXTE data}

We have also used the HEXTE standard mode data. The data were
corrected for background measurement duty cycle and for the $40 \%$ to
$60\%$ deadtime effects (using version 0.0.1 of hxtdead). Once again,
using light curves, color-color, and hardness-intensity diagram
analysis, no spectral or intensity variation could be
inferred. Therefore, we have combined the mean spectra of the two
HEXTE clusters (using addspec 1.2.0). Response matrices have been
downloaded from the UCSD web site\footnote
{http://mamacass.ucsd.edu:8080/hexte/hexte$_-$calib.html}.

\begin{figure}[t]
\psfig{figure=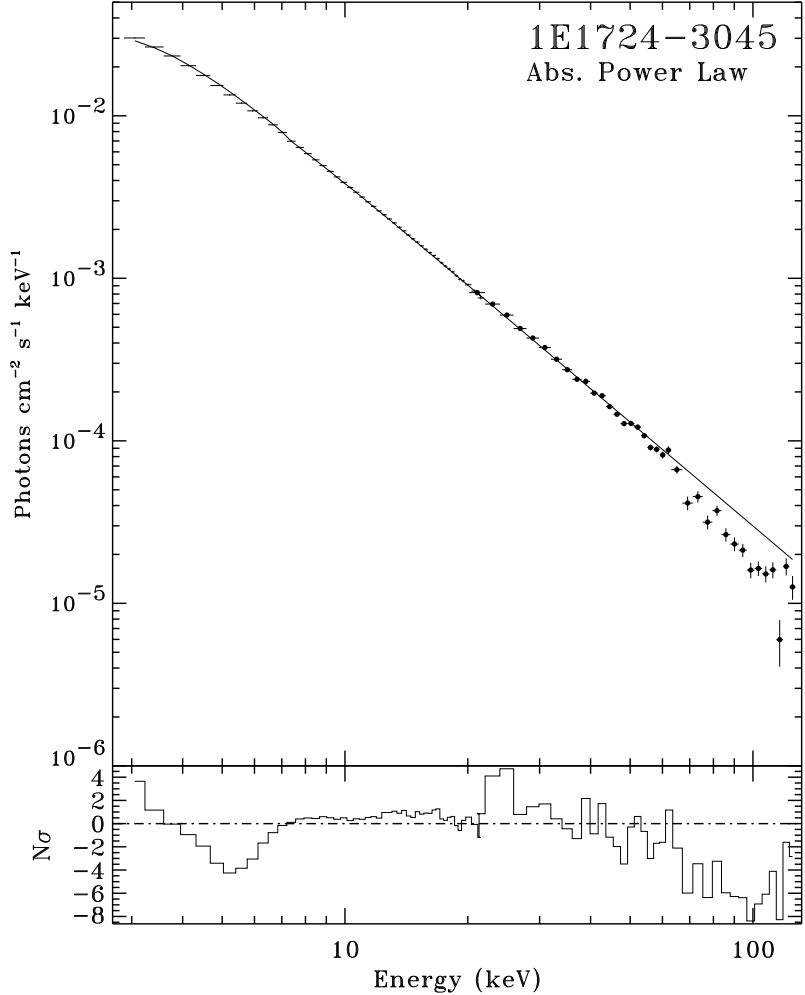,height=14cm}
\caption{Unfolded combined PCA and HEXTE spectra 
of \onee~for a power law fit. A clear spectral break occurs 
around 50 keV and the fit is not satisfactory below 10 keV.}
\label{plfit}
\end{figure}

\section{Spectral analysis of \tz} 

For spectral fitting, we have used XSPEC version 10.0. When fitting
simultaneously PCA and HEXTE spectra, we found that the optimum energy
ranges for the fit were 3--20 keV for the PCA and 20--150 keV for
HEXTE. In a combined fit, the relative normalization between the PCA
and HEXTE spectra has been left as a free parameter of the models.

\subsection{Fit by a power law}

An absorbed power law is the simplest model we have tested (see
results in table 1).  The reduced chi-squared of the fit is \chisqr
$\sim 5.1$ for 82 dof. Figure \ref{plfit} shows why the fit is not
satisfactory. At low energy, the spectrum is flatter. If this
flatening goes down to 2.5 keV, the residuals suggest the presence of
a soft excess. At higher energy a clear cutoff or break in the HEXTE
spectrum around 50-60 keV appears. On the other hand, the column
density (3.7 \nhv) is significantly larger that what has been
previously observed with ASCA and SAX.  These results are suggestive
of the following features : (1) a spectral break in the HEXTE range,
(2) a harder spectrum below say 10 keV, (3) a possible soft excess
below 5 keV.

For comparison with SIGMA data, it was interesting to fit the HEXTE
data alone in the nominal SIGMA range of 35--150 keV with a single
power law model. The index we derived was $2.7\pm0.1$ consistent with
the SIGMA mean value ($3.0 \pm 0.3$). However the power law fit was not
satisfactory (\chisqr = 1.84 for 61 dof) and the $\sim 50$ keV break was
readily visible, even in this restricted energy range.

\subsection{Fit with a Comptonized model and a Blackbody}

In the light of recent observations of the source with ASCA and SAX,
we have fit the broad band spectrum with a recently developped
Comptonization model (the so-called ``comptt'' in XSPEC) which can
produce spectra with a high energy cutoff and a flatenning below few
keV. Both characteristics were observed in the \tz~spectra. The
Comptonization parameters are: the electron temperature \kte, optical
depth of the electron cloud $\tau$ and a temperature of the ``seed''
photons of
\ktw~assuming a Wien-type distribution (Titarchuk, 1994). This model
inputs two different geometries: disk and sphere. Despite these
interesting properties, it is clear that the ``Comptt'' model alone
could not take into account the low energy soft excess (below 5
keV). So, we have tried to fit our spectra with a composite model of a
blackbody plus a Comptonized spectrum (see results in table 1).

Adding a blackbody component makes impossible to leave the
\nh~as a free parameter because of the strong correlation existing
with this parameter and the amplitude of the blackbody. Furthermore,
the points constraining their values are restricted to a narrow energy
range. Consequently, the \nh~was fixed to its nominal ASCA value of
1.0 \nhv (Barret \etal, 1998).

\begin{figure}[t]
\psfig{figure=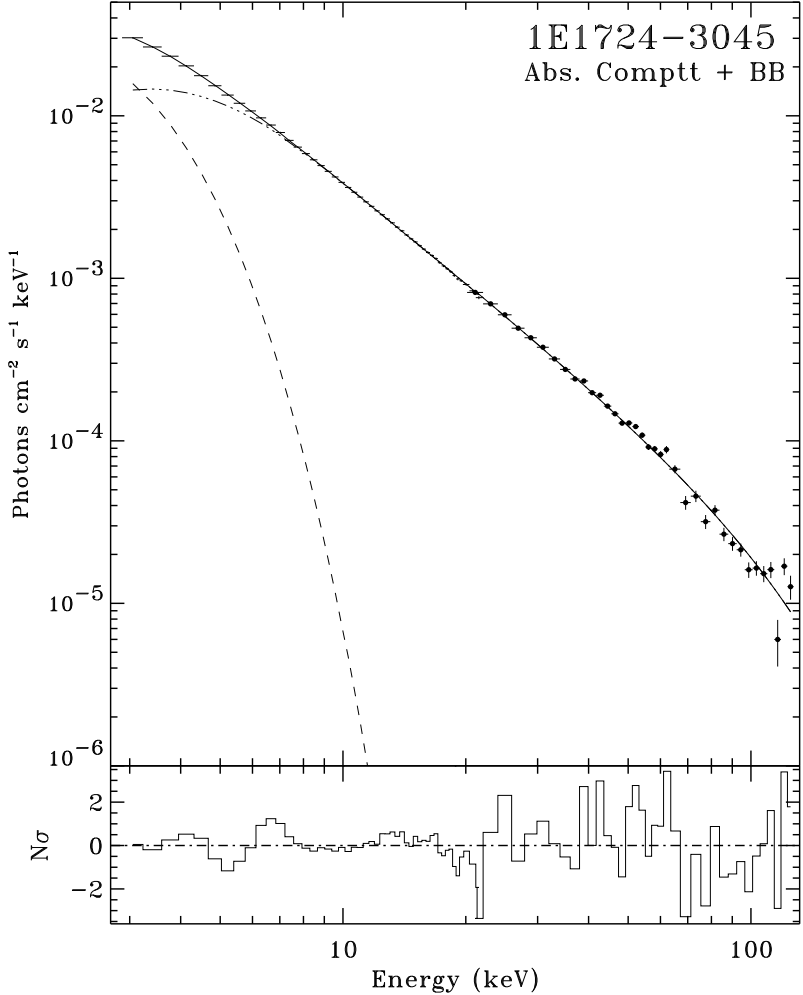,height=14cm}
\caption{The PCA and HEXTE combined 
unfolded spectrum for the Comptonization model presented in Titarchuk
(1994). The high energy cutoff is nicely fitted. A blackbody component
has been added to the fit to account for the soft excess below 5
keV.}
\label{compttbb}
\end{figure}

Figure \ref{compttbb} shows the unfolded spectrum and residuals. It
shows that the wave shape below 5 keV has been removed while the fit
has accomodated an acceptable \chisqr value (See table 1). Using a disk
instead of a sphere geometry yields essentially the same value for
\kte~and \ktw; the main effect being on the optical depth which
decreases from $\sim 3$ to $\sim 1.$

\begin{table}[t]
\begin{center}
\begin{tabular}{|llc|}
\hline 
\hline 
Model & Parameters &     \\ \hline
          & \nh               &  $3.7 \pm 0.1 $      \\
Power Law & $\alpha$          &  $2.1 \pm 0.05$      \\
          & \chisqr (d.o.f)   &  5.12 (82 dof.)      \\
\hline
          & \nh               &  $1.0$ (fixed)       \\
          & kT$_{BB}$         &  $0.67 \pm 0.07$ keV \\
          & kT$_w$            &  $1.19 \pm 0.1$ keV  \\
BB+Comptt & kT$_e$            &  $29.2 \pm 2$ keV    \\
          & $\tau$            &  $2.9 \pm 0.2$       \\
          & $\chi^2$ (d.o.f)  &  1.10 (79 dof.)      \\
\hline
\hline
\end{tabular}
\caption{Best fit spectral parameters 
(Comptt=Comtonization model from Titarchuk (1994),
BB=Black body). Error are give at the 90$\%$ confidence level for
variation of one single parameter.}
\end{center}
\label{fitresu}
\end{table}





\section{Discussion}

\onee~belongs to the class of bright and persistent LMXBs located in
globular clusters. We observed this ``Atoll'' source in its hard
state, during which it emits hard X-rays. The source did not show any
spectral variability. The RXTE spectrum could be well described by a
hard Comptonized component (\kte $\sim 30$ keV, \ktw=1.2 keV, $\tau \sim
3$), plus a soft component which could be fit by a Blackbody
(\ktbb=0.7 keV,
\rbb=10-11 km). The origin of the soft component is unclear (neutron
star surface or optically thick boundary layer). The fact that our
RXTE spectrum is consistent with the SAX LECS-MECS (Guainazzi \etal,
1998) and ASCA (Barret \etal, 1998) ones suggests that this source
spends most of its time in such a hard state. 

Barret, McClintock and Grindlay (1996) proposed that X-ray bursters
(XRBs) can be distinguished in two ways from black hole candidates
(BHCs) using luminosity criteria.  First, no XRBs seem to be able to
emit hard X-ray tails when their 1-20 keV luminosity is larger than
$\sim 2 \times 10^{37}$ \ergs.  During our observation, the 1-20 keV
luminosity was 1.0 $\times 10^{37}$ \ergs~at 7.7 kpc. So the hard
X-ray emission is indeed associated with a low X-ray intensity state.
Second, no XRBs have hard tails brighter than $\sim 2 \times 10^{37}$
\ergs.  Again, this criterion is satisfied for \onee~ as the 20-200
keV luminosity was 6.4 $\times 10^{36}$ \ergs.

Recently it has been suggested that the Comptonization temperature
could also be used as a criterion to discriminate between BHCs and
XRBs: the former would have temperature \kte~$\ga$~50 keV, whereas for
the latter it would be $\la 30$ keV (Zdziarski \etal, 1998). Our
result fits in this scheme.

\section{Conclusions}

It is already well known that ``Atoll'' sources have timing properties
similar to BHCs (Van der Klis, 1995, see also Olive et al. 1998 for
\onee). The present result demonstrates that they also share
similarities with respect to their spectral properties. Both systems
emit hard X-rays and Comptonization seems to be the dominant emission
mechanism. Further broad band observations are needed to test the
luminosity criteria over a larger sample of sources, and to definitely
assess whether BHCs and XRBs can indeed be distinguished on the basis
of the electron temperature of the Comptonizing cloud.

\section{References}
\vspace{-5mm}
\begin{itemize}
\setlength{\itemindent}{-8mm}
\setlength{\itemsep}{-1mm}

\item [] {Barret}, D., \etal, 1991,  {\em ApJL}, {\bf 379}, L21

\item [] {Barret}, D., {Mc Clintock}, J. E. and  {Grindlay}, J. E., 1996,
{\em ApJ}, {\bf 473}, 963

\item [] {Barret}, D., \etal, 1998,  {\em  A\&A}, in press

\item [] {Bradt}, H.~V., Rothschild, R.~E. and Swank, J.~H., 1993, {\em A\&A Supp. Series}, {\bf 97}, 355

\item [] {Churazov}, E., et al., 1997, {\em Advances in Space Research}, {\bf v. 19 Issue 1}, 55

\item [] {Goldwurm}, A., et al., 1995, {\em Advances in Space Research}, {\bf v. 15 Issue 5}, 41

\item [] {Guainazzi}, M., \etal, 1998, {\em A\&A}, {\bf 339}, 802

\item [] {Olive}, J.~F., \etal, 1998, {\em A\&A}, {\bf 333}, 942

\item [] {Swank}, J.~H., \etal, 1977, {\em ApJL}, {\bf 212}, L73

\item [] {Tavani}, M. and {Barret}, D., 1997, in D.C.D., M. {Strickman}, and J.~D. {Kurfess} 
(eds.), {\em Proc. of the 4th Compton symp.}, AIP Conf. Proc., {\bf 410}, 75

\item [] {Titarchuk}, L., 1994, {\em ApJ}, {\bf 434}, 570

\item [] {Van der Klis}, M., 1995, in W.~H.~G. {Lewin}, J. {Van Paradijs}, and E.~P.~J. {Van Den Heuvel} (eds.), {X-Ray Binaries},{\bf  26}, 252, Cambridge University Press

\item [] {Zdziarski}, A.~A., \etal, 1998, {\em Mon. Not. R. Astron. Soc.}, in press

\end{itemize}

\end{document}